\documentclass[journal]{IEEEtran}
\usepackage{cite}

\ifCLASSINFOpdf
   \usepackage[pdftex]{graphicx}
\else
\fi
\usepackage{amsmath}
\usepackage{amssymb}

\usepackage{url}

\usepackage{txfonts}
\usepackage{xcolor}

\newcommand{\Y}[1]{{ \color{blue} {\bf} #1}}

\usepackage{CJKutf8}
\usepackage{graphicx}

\begin{document}
%
%
%

\title{Simulation of the Space-Charge Limited Current Density for Time-Variant Pulsed Injection}

%
\begin{CJK}{UTF8}{gbsn}
\author{H.~Huang,~\IEEEmembership{Member,~IEEE,}
        and Y. Liu (刘泱杰)\IEEEauthorrefmark{1},~\IEEEmembership{Member,~IEEE}
\thanks{H. H. and Y. L. are presently with Faculty of Physics and Electronic Science, Hubei University, 430062 Wuchang District, Wuhan, P. R. China.  

 \IEEEauthorrefmark{1} Corresponding author, yangjie@hubu.edu.cn.}
\thanks{Y. L. was with Computational Nano-Electronics Laboratory, Division of Micro-Electronics, School of Electrical and Electronic Engineering, 50 Nanyang Avenue, Nanyang Technological University, 639798 Singapore.}

\thanks{Manuscript received July 25, 2018; revised 19 November, resubmitted 7 December 2018; re-revised 10 February 2019, re-submitted 11 February 2019 (TPS11450.R2). H. H. is supported by the Natural Science Foundation of Hubei Province of China (2018CFC797). L. Y. acknowledges financial support of his PhD stipend from Tier II Academic Research Fund (MOE2008-T2-01-033) and Antennas Group travel grant (EECSRC2/4a). }}

%
%

\markboth{Journal of \LaTeX\ Class Files,~Vol.~xx, No.~x, August~2018}%
{Huang \MakeLowercase{\textit{et al.}}: Bare Demo of IEEEtran.cls for IEEE Journals}
%



\maketitle

\end{CJK}

\begin{abstract}
Space charge limited (SCL) current density for time-invariant injection via the diode cathode is the maximum transportable density, while it can be leveraged higher when the injection pulse length is shorter than the transmit time for electrons (i.e. under \emph{short-pulse} condition). 
However, both known limits apply for the time-invariant injection condition and the role of time-varying current density for injection remains elusive. In this paper, we numerically investigate the space-charge limited electron flow with time-variant injection. Using particle-in-cell simulation, four different time-variant profiles for electron injection are enforced, and the maximum current densities are determined resulting from the space charge effect for various pulse lengths. We speculate that time-variant density of injection via the diode cathode will contribute to transport enhancement. 

\end{abstract}


%


\section{Introduction}
\begin{figure}[!htbp]
\centering
\includegraphics[width =0.45\textwidth, keepaspectratio]{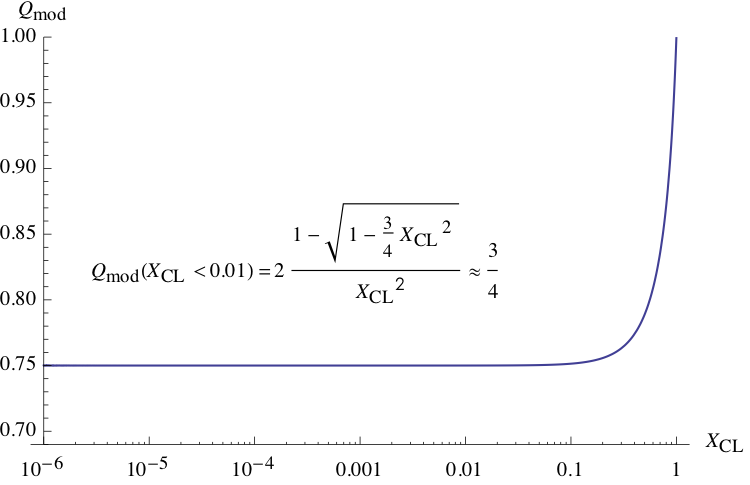}
\caption{\label{fig:Qmod} 
(Color online) The ratio of transported charge between the long pulse limit and short pulse $Q_{\rm mod}$ as a function of the normalised pulse length $X_{\rm CL}$ = $\tau_p / \tau_{\rm CL}\leq 1$. }

\label{fig_sim}
\end{figure}

%
%
%
%
\IEEEPARstart{S}{pace-charge-limited} (SCL) current density describes the maximum current density transportable for a diode under a voltage drop between its cathode and anode, when the injection current density from the cathode is time-invariant.  For a vacuum diode with a gap spacing of $D$ and a dc voltage of $V_g$ , it is known as one-dimensional (1D) Child-Langmuir (CL) law \cite{Child1911, Langmuir1913}, given by
\begin{equation}\label{JCL}
J_{\rm CL}=\frac{4\epsilon_0}{9D^2}\sqrt{\frac{2e}{m_0}}V_g^{3/2}, 
\end{equation}
where $\epsilon_0$, $e$ and $m_0$ are respectively, vacuum permittivity, electron charge and electron mass. 
Space charge limited electron flow in one-dimensional case occurs if the charge of the emitted electrons is sufficient to suppress the cathode electric field to zero, when the electrostatic potential distribution function and charge density are respectively
\begin{eqnarray}\label{Vxdis}
\phi(x)&=&V_g\left(\frac{x}{D}\right)^{4/3}, \nonumber\\
\rho(x)&=&\frac{4}{9}\epsilon_0 V_{\rm g}\left({D^2x}\right)^{-2/3},
\end{eqnarray}
where $x$ is the spatial coordinate. This 1D classical CL law above assumes the long-pulse condition, i.e. constant injection current density from cathode, and extends to the short pulse model when the injection pulse length is less than the transmit time of electrons~\cite{Valfells2002}. The short pulse model pushes the maximum transportable current density higher than the CL law predicts. Therein the short pulse SCL current density, i.e. Valfells' formula, can be expressed as
(cf. black line of Fig.~\ref{fig:abs})
\begin{equation}\label{Valfells}
J_{\rm S}(X_{\rm CL})=2\frac{1-\sqrt{1-{3}X_{\rm CL}^2/4}}{X_{\rm CL}^3} J_{\rm CL},
\end{equation}
where $X_{\rm CL}$ = $\tau_p / \tau_{\rm CL}\leq 1$ is the normalized pulse length and 
$\tau_{\rm CL}= 3D\sqrt{m_0/{2eV_g}}$ is the CL transit time to traverse the diode. Physically, a bunch of electrons is injected into the gap with a constant current density $J_{\rm S}$ over length $\tau_{\rm p}$ ($0 \leq t \leq \tau_{\rm p}$). This equivalent (virtual) diode model captures the physical feature of virtual cathode formation localised in time and space~\cite{Valfells2002, LiangP2018, Campanell2018} and results in Eq.~\eqref{Valfells} above~\cite{Valfells2002}. Furthermore, the ratio of transported charge between the long pulse limit and short pulse, can be written as
\begin{equation}
Q_{\rm mod}(X_{\rm CL})=\frac{J_{\rm S}\tau_{\rm p}}{J_{\rm CL}\tau_{\rm CL}}=2\frac{1-\sqrt{1-3X_{\rm CL}^2/4}}{X_{\rm CL}^2},
\end{equation}
after substituting Eqs.~\eqref{JCL} and \eqref{Valfells}~\footnote{An interesting note for CL electron flow is $J_{\rm CL}\tau_{\rm CL}=4\epsilon_0V_g/3D\sim V_g/D$. }. Thus the total transported charge cannot exceed that under convectional CL condition and in short pulse limit, the ratio $Q_{\rm mod}\simeq 3/4$, as indicated in Fig.~\ref{fig:Qmod}~\cite{MyThesis}. The critical current density Eq.~\eqref{Valfells} for a short-pulse electron flow, results from considering only virtual cathode beam front and sidestepping internal repulsion force within the pulse spatially.

 




%
%

Although a substantial amount of investigations on SCL effect have been performed~\cite{Lau91, Luginsland1996, Lau01, Koh2005, Luginsland2002, Ums01, Rok2003, Pedersen2010, ZhangP2017}, most of them assume injection of a time-uniform current density. Some subsequent progress, however, has been made to understand the time-variance effect of an injection current density~\cite{Griswold2010, Zhu2011, Griswold2012, Caflisch2012, Griswold2016}. It was even speculated that there might be an upper bound of time-averaged SCL current density for a given time-dependent current, which could be higher than the time-independent case such as CL law. However, the exact such time profile for injection current remains unknown~\cite{YL2015}. Furthermore, the long-pulse time-variant injection was already investigated and believed to play little role to enhance the SCL density~\cite{Liu2017}. Note that our numerical code gives more accurate prediction for relativistic SCL density other than Jory's analytic formula (Fig.~1 in \cite{Jory1969}) by considering time-variance of injection pulse numerically~\cite{Liu2017}. When the injection current density becomes pulsed and time-varied, the internal repulsion within the electron beam will reshape its spatial profile into nonuniform in space. Then Valfells' formula~\cite{Valfells2002} ceases to apply, and numeric methods are then used to address the transient behaviour for the short-pulse electron flow. \Y{The time-varying issue for pulsed injection should be important in the sense of reaching SCL by tuning short pulse plasma source in time in a diode emission setup experiment, such as a photoinjector~\cite{Valfells2002}. As the ultrafast lasers up to femtosecond range and nano-fabricated samples become popular in modern experimental setup of plasma emission, it is curious to look at the space-charge effect down to the nano spacial and ultrafast temporal scales. It is also noted that we intensionally neglect other emission mechanisms (field emission etc.) in order to reveal the role time-varying injection plays in SCL electron transport. }

In this paper, we are motivated to investigate the characteristics of different short time profiles in order to enhance time-average current density as compared to the short-pulse Valfells' formula \cite{Valfells2002}. 
In particular, the electrons are injected with negligible initial velocity into the gap following four prescribed time profiles, and the particle-in-cell (PIC) simulation in one-dimensional limit is performed to obtain the time-average SCL current density. It will be revealed that the time-average SCL current density of such time-varying profiles can be higher by a factor of 2 to 3 as compared to time-invariant injection. 
Discussions on the time dependence of the electric field on cathode and anode are also given.

\section{Model and simulation}
In this section, our problem of short-pulse electron flow in a 1D classical diode is described and its simulation set up for PIC simulation. 
Consider a gap of spacing ($0\leq x\leq D$) \Y{on which an external fixed electric field $F_{\rm dc} = V_g / D$ is imposed.}
The electrons are injected from the cathode $x= 0$ with a time-variant current density $J(t)$ over a pulse length 
$\tau_p \leq \tau_{\rm CL} = 3D\sqrt{m_0/{2eV_g}}$. The emitted electrons are then exerted only by the direct electric field within the spacing, instead of by other emission mechanism~\cite{Barwick2007, Yalunin2011, Pant2012, Wendelen2012, Liu2014}. 
Here, it is assumed that $J(t)$ can be expressed as
\begin{equation}\label{Jt}
J(t)= \beta_m \cdot  j_m(t),
\end{equation}
where $\beta_m$ is the magnitude of the current density, and time-profile $j_m(0 \leq \bar{t}=t/\tau_p \leq 1)$ is a particular time-dependent profile assumed in the model (for $m\in\mathbb N, t\in[0,\tau_{\rm p}]$) among (see Fig.~\ref{fig:jit} below)
\begin{eqnarray}\label{jt}
j_0(\bar t)&=&1,\\
\label{jt1}j_1(\bar t)&=& 2 \bar{t},\\ 
\label{jt2}j_2(\bar t)&=&2 (1-\bar{t}),\\
\label{jt3}j_3(\bar t)&=&H\Bigl(\frac{1}{2}-\bar{t}\Bigr)\cdot 4 \bar{t}+H\Bigl(\bar{t}-\frac{1}{2}\Bigr)\cdot 4(1-\bar{t}),\\
\label{jt4}j_4(\bar t)&=& 3\bar{t}^2.
\end{eqnarray}
Here, $j_m(\bar{t})$ has been normalized ($\int_0^1 j_m(\bar{t}) {\rm d}\bar{t} = 1$), and $H(\cdot)$ stands for Heaviside step function. \Y{Such time profiles for injection pulses are assumed in the hope to reveal some dynamic feature of short-pulsed plasma transport. Polynomial functions are chosen for they serve as the basic functions in Taylor expansion for a time-variant function in general.  } Accordingly, the whole electron count number density (per unit emitter area) $N_{\rm EC}$ is defined as
\begin{equation}\label{NEC}
N_{\rm EC}=\int_0^{\tau_{\rm p}}\frac{J(t)}e{\rm d}t.
\end{equation}

\begin{figure}[h]
  \centering
	\includegraphics[width=0.45\textwidth,keepaspectratio]{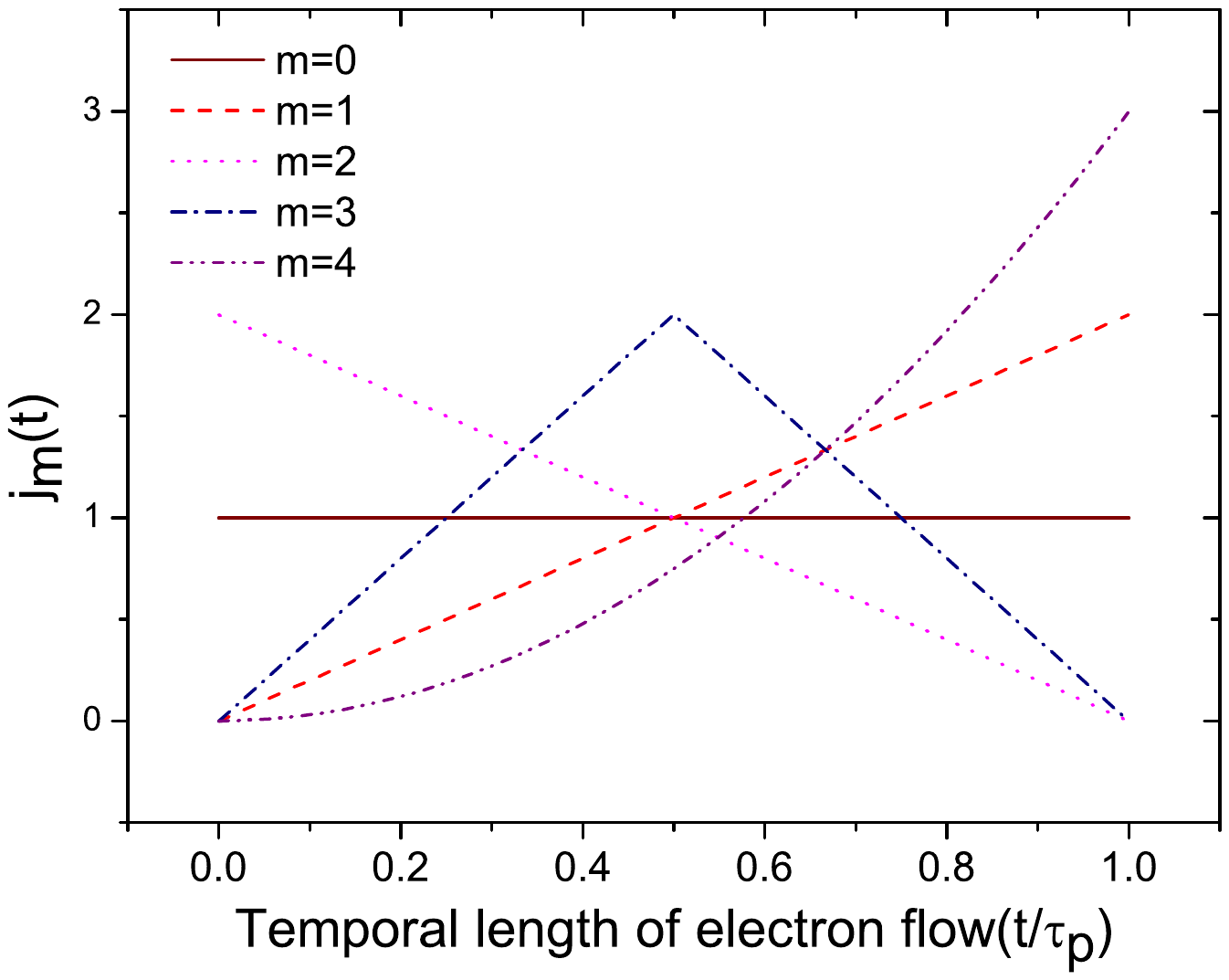}
\caption{\label{fig:jit}
(Colour online) The five normalized time profiles $j_m(t)\vert_{m = 0, 1,2,3, 4}$ assumed in the model [see Eq.~\eqref{Jt}] to be used in PIC simulation.}
\end{figure}


To check whether any representative of the time-varying current densities $J_m(t)\vert_{m>0}$ is able to transcend in average the 
constant profile $m=0$ case in Eq.~\eqref{Jt}, a commercial PIC software (\texttt{VORPAL4.2})~\cite{Nieter2004,VORPAL4p2} is used to
determine the occurrence of SCL current at the end of the short pulse.
The simulation is constructed in one-dimension limit in order to unfold the basic phenomenon for space-charge-limited plasma flow.  
The SCL current is determined by increasing the magnitude of the injected current $\beta_m$ until electron reversal is observed ~\cite{Mahalingam2009, Chen2011, Liu2012b}. 

We also note here the procedures to set up \texttt{VORPAL} simulation to explore the SCL short pulse flow. As mentioned in order to avoid additional problems of plasma expansion in multidimensional space(as 2D or 3D cases), computation is performed in one-dimensional limit to study the \Y{very} effect of time-variant electron flow. Then global uniform grid ({\itshape uniCartGrid} in Python script) is adopted to divide equally 1D space and assign values of electro-static potentials for both walls(cathode and anode) as Dirichlet boundary conditions. For any particle emitted within simulation grid~\cite{VORPAL4p2}, a particle source is defined according to physical properties of electrons. A bunch of electrons are then emitted at rest from cathode surface (eg. $x=0$ {\itshape emitSurface} in script writing). Time-variant injection density in Eqs.~\eqref{jt}-(10) is also included in this part to vary electron number upon time emitted from cathode. The space-charge limit is reached if until the end of short-pulse flow, zero or negative momentum of electrons within the diode gaping is observed from VORPAL data. To avoid the trouble of particles running out of the simulation grid, two particle-absorbers adjacent to both ends of grid are placed to absorb any particle diffused across the demarcation borders. Our procedure to emit electrons from cathode is calculated from the perspective of current density rather than of particle number density in space. This emit-flux-mechanism avoids indirect conversion from current density to particle density.  

\section{Results and discussion}\label{sec:results}

\begin{figure}[htb]
  \centering
	\includegraphics[width=0.5\textwidth,keepaspectratio]{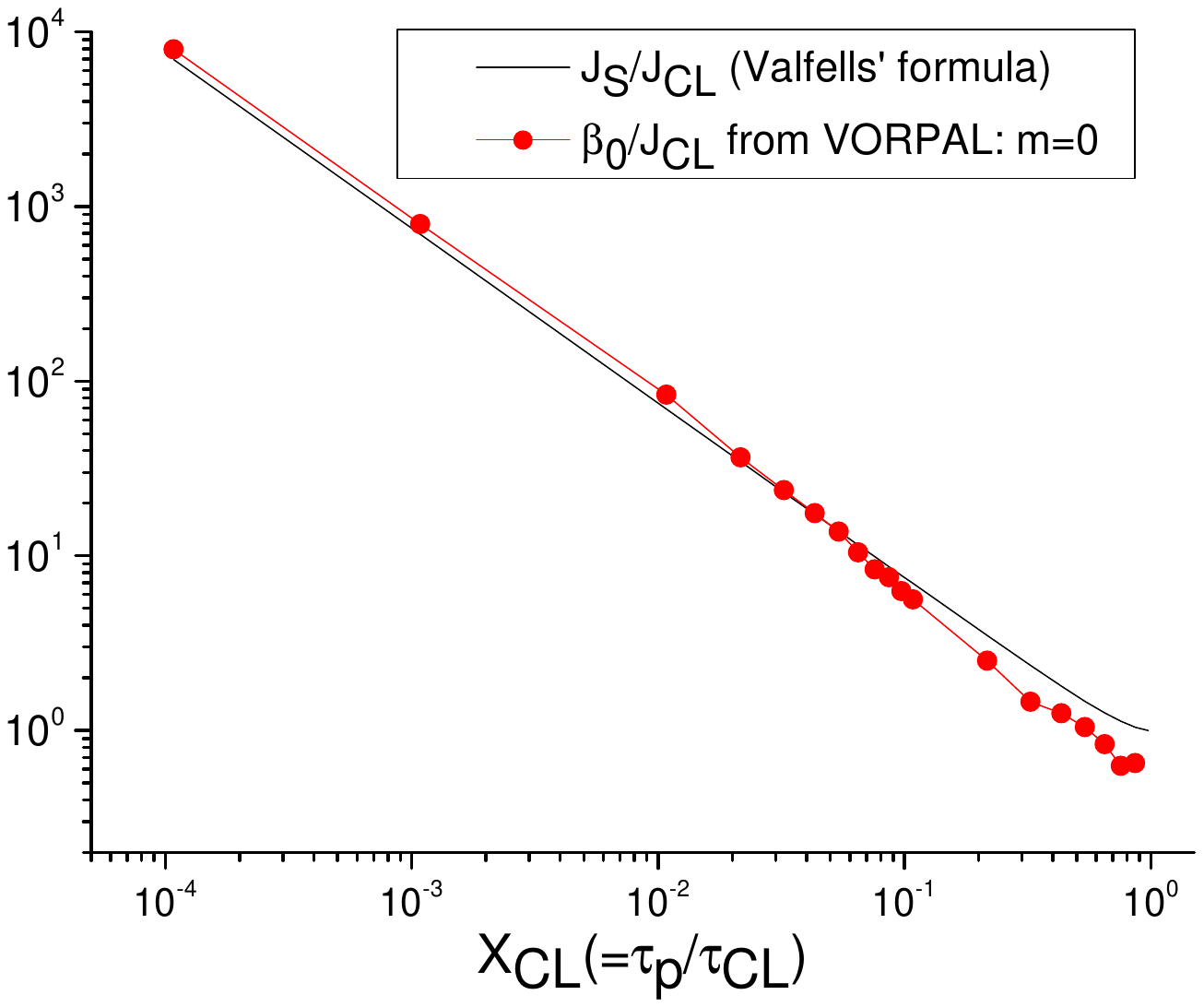}
\caption{\label{fig:abs} 
(Color online) Normalised SCL density with respect to $J_{\rm CL}$ for short pulse condition: $\beta_0/J_{\rm CL}$ (symbol) and $J_{\rm S}/J_{\rm CL}$ (line), plotted as a function of normalized pulse length $X_{\rm CL}$. The parameters are $D = 1 {\rm cm}, V_{\rm g}= 30 {\rm MV}$, and SCL transit time $\tau_{\rm p} = 9.24 {\rm ps}$. The reason why a low value occurs near 10{\rm ps} is explained near the end of Section~\ref{sec:results}. 
 }
\end{figure}

We shall present our simulation results for the SCL short flow and give some discussion in this section. For the reference case, i.e. time-invariant injection density at $m = 0$ in Eq.~\eqref{jt}, determines the SCL current density as a constant $J_0 = \beta_0$. Thus it is expected that this time-invariant case $m=0$ should recover Valfells' formula Eq.~\eqref{Valfells}, which means $\beta_0=J_{\rm S}$, as shown in Fig.~\ref{fig:abs}.

In Fig.~\ref{fig:abs}, the SCL average densities are compared [normalised to Valfells' formula Eq.~\eqref{Valfells}] between $\beta_0$ ($m$ = 0 case) and $J_{\rm S}$ at $D = 1 {\rm cm}, F_{\rm dc} = 3 {\rm GV/m}$ ($V_g$ = 30 MV) as a function of $X_{\rm CL}$ up to $\lesssim1$, where the electron transit time is $\tau_{\rm p}= 9.24{\rm ps}$.
The comparison shows good agreement with $\beta_0$ though slightly higher than those predicted by Eq. \eqref{Valfells} at small $X_{\rm CL}$. 

\begin{figure}[htbp]
  \centering
	\includegraphics[width=0.53\textwidth,keepaspectratio]{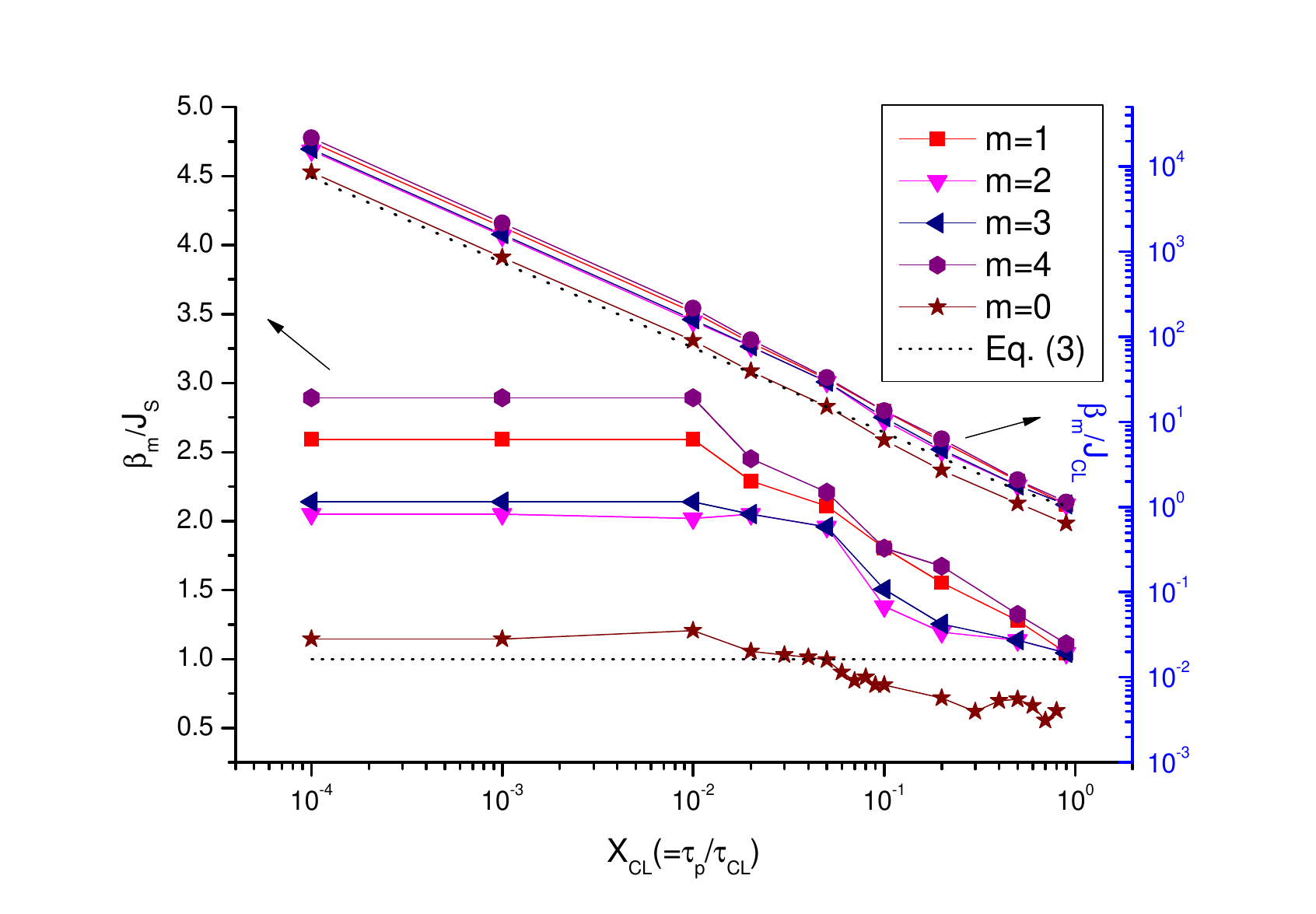}
\caption{\label{fig:result}
(Color online) The magnitude of current density $\beta_m$ ($m$ = 0, 1, 2, 3, 4) with repect to $J_{\rm S}$(left ordinate) and $J_{\rm  CL}$ (right ordinate)
from Eq.~\eqref{Valfells}, as a function of normalized pulse length $X_{\rm CL} < 1$.
The same parameters are used as in Fig.~\ref{fig:abs}. Note that the dashed line indicates the unity level. 
}
\end{figure}

As shown in Fig.~\ref{fig:result}, the determined values of $\beta_m$ (for $m$ = 0, 1, 2, 3 and 4) are plotted in terms of $J_{\rm CL}$ and $J_{\rm S}$ as a function of normalised pulse length $X_{\rm CL}$. The ratios $\beta_m/J_{\rm CL}$ in logarithm show a clear double-logarithm scaling with pulse length $X_{\rm CL}$. The ratios $\beta_m/J_{\rm S}$ demonstrate that the highest enhancement is achieved in $m= 4$ case, followed by $m=$1 , 3, 2 and 0.
The data indicate that an increasing time-varying of $J(t)$ will have a higher time-average current magnitude
($\beta_4 > \beta_1 > \beta_3 > \beta_2$).
For a linearly increasing function $m$ = 1 case, the enhancement is about 2.6.
For a linearly decreasing function $m = 2$ case, it is about 2. 
In all cases, the enhancement is above 2 for small $X_{\rm CL} < 0.01$, and the enhancement deceases with large $X_{\rm CL}$ approaching 1. The reduction to unity is attributed to that the longer pulse length inhibits the electron pulse tail from entering diode before the pulse front terminates at anode.

\begin{figure}[htbp]
  \centering
	\includegraphics[width=0.54\textwidth, keepaspectratio]{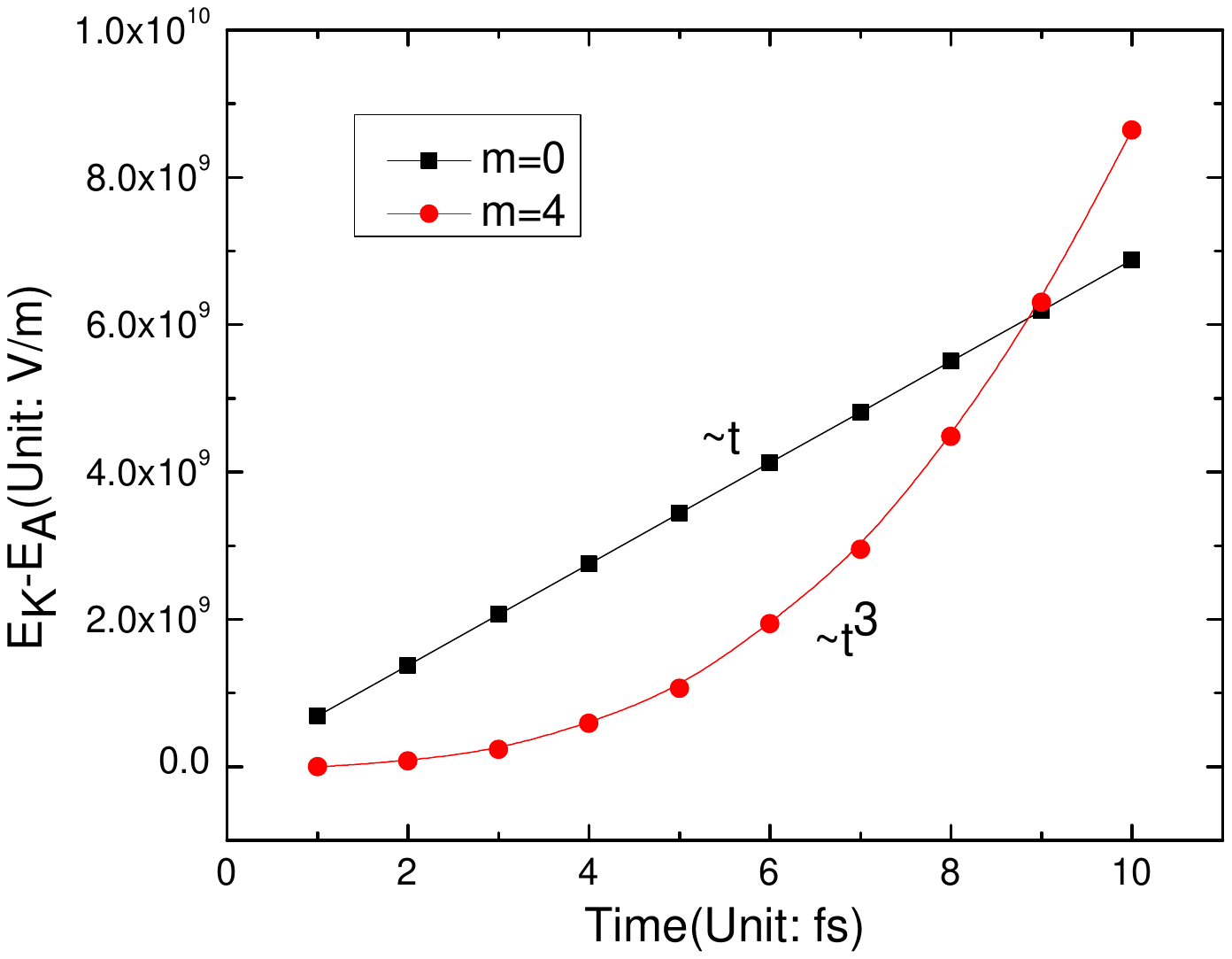}
\caption{\label{fig:Et}
(Color online)The time dependence of the difference between the electric field on the cathode and anode, $E_{\rm K}(t)-  E_{\rm A}(t)$ for $m$ =0 case:  (black square) and $m$ = 4 case (red dot), which shows, respectively, $t$ and $t^3$ scaling.
The used parameters are $D$ = 1 cm, $V_{\rm g}$ = 30 MV, and $\tau_{\rm p} =10 {\rm fs}$.
}
\end{figure}

To affirm the effect of time-variance of electron flow in simulation,  electric fields of cathode and anode $E_{\rm K}(t)$ and $E_{\rm A}(t)$ are retrieved respectively, the difference of which is plotted in Fig.~\ref{fig:Et}. One can interpret $E_{\rm K}(t)-  E_{\rm A}(t) $ as follows. According to 1D Gauss's law and conservation of charge, the difference of the electric field between the cathode and anode  is proportional to integral of current density $J(t)$, given by
\begin{eqnarray} 
\label{Ediff}
\label{EK-EA}E_{\rm K}(t)-  E_{\rm A}(t)=-\int_0^D\frac{\partial E(x)}{\partial x}{\rm d}x\nonumber\\
=\frac{1}{\epsilon_0}\int_0^D\rho(x){\rm d}x=\frac{1}{\epsilon_0}\int_0^t J(t'){\rm d}t',
\end{eqnarray}    
where $\rho(x)$ the charge density without its sign, and $t'$ a dummy variable. In Fig.~\ref{fig:Et}, the field difference between the cathode and anode for two cases $m = 0$ and  4 are presented, which indicates that [$E_{\rm K}(t)-  E_{\rm A}(t)] \sim t$ (for $m$ = 0 case) and $t^3$ (for $m$ = 4 case) as expected from Eq. \eqref{Ediff}. Then our simulation is assured to include the time-dependence for injection flow. 

\begin{figure}[htbp]
  \centering
	\includegraphics[width=0.47\textwidth, keepaspectratio]{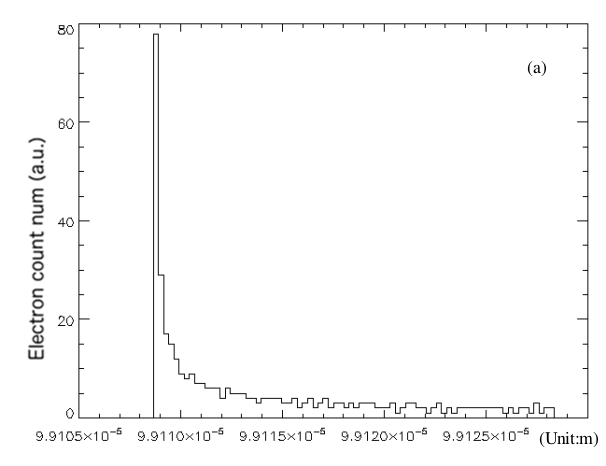}\\
	\includegraphics[width=0.47\textwidth, keepaspectratio]{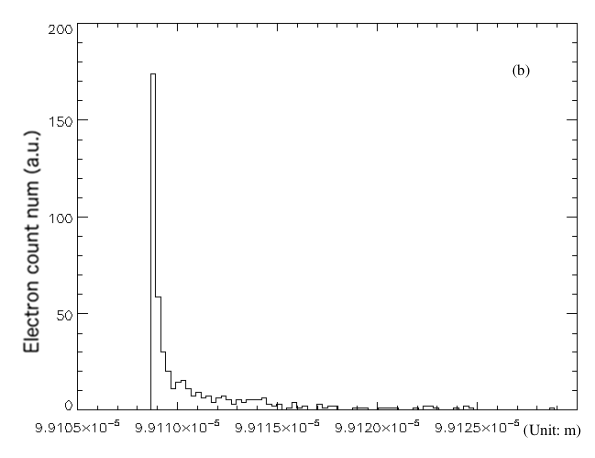}
\caption{\label{fig:electron-num}
Histograms of electron count number (arbitrary unit) versus $x\in[0, D/100]$ space between cathode and anode at the pulse end of short electron flow ($t=\tau_{\rm p}$), for (a) $m$ = 0 and (b) $m$ = 4 case. Produced from \texttt{VorpalView 4.0.0}~\cite{VORPAL4p2} and \texttt{Mathematica 9}~\cite{MM9}. The same parameters as in Fig.~\ref{fig:Et} are used. Reproduced from ~\cite{MyThesis}.  
}
\end{figure}

In Fig.~\ref{fig:electron-num}, the histograms for the electron counts are shown inside the gap near cathode at the end of the pulse ($x\in [0, D/100], t = \tau_p$) for (a) $m= 0$ and (b) $m = 4$ case, as the latter gives the highest average current density transported. For injection of case $m = 4$, the electron distribution is more localised near cathode as compared to case $m = 0$. The reason why it transports more current is that in $m=4$ case most part of the electron flow is squeezed in the pulse end, although further investigation is still required to affirm our speculation. 

Finally, all data in Figs.~\ref{fig:abs} and ~\ref{fig:result} are valid for only for short pulse regime $X_{\rm CL} < 1$ where the electron pulse length is less than the transit time. Our simulation shows that for presented data electron flow spends $9.24{\rm ps}$ to transmit the space. For $X_{\rm CL} \geq 1$, the front of electron pulse will be collected by the anode while its trail has not yet emitted from the cathode, which means PIC simulation fails to check the genuine physical limit of space-charge. Thus the determined transported current density de facto reduces to the steady one--CL limit in Eq.~\eqref{JCL}.

It is concluded that time-variance of injection current density can enhance the SCL current density from our simulation data, though our chosen four time-varying profiles are far from exhaustive and more investigation is required~\footnote{See further simulation for more time-profiles in Ch. 3 of the PhD dissertation~\cite{MyThesis}. }.

\section{conclusion}

In this paper, we present numerical results using PIC simulation to study the effect of time-varying injection for a pulsed electron flow into diode to reach the space charge limited (SCL) condition. It is shown that time-varying injection profile can contribute to obtain a higher time-average SCL current density, than the time-invariant short pulse SCL one~\cite{Valfells2002}, although further detailed study is still required to pinpoint the issue. Time-dependence should be in favor of leveraging space-charge-limit for short-pulse injection. These results may pave the way in using laser or other mechanism to excite a specific time-varying injection at the high-current regime.


%



\section*{Acknowledgment}

The \texttt{VORPAL} simulation was performed on computing server Axle of IHPC, A*STAR in Singapore. L. Y. would like to thank {\'A}g{\'u}st Valfells, Koh Wee Shing and Christine Roark for their helpful discussion and persistent pedagogy, and South China Normal University for hosting his 2013 visit.

\ifCLASSOPTIONcaptionsoff
  \newpage
\fi



\bibliographystyle{IEEEtran}
%



\bibliography{bib5}

%

\begin{IEEEbiography}{Hao Huang}
Biography text here.
\end{IEEEbiography}

\begin{IEEEbiography}{Yangji\'e Liu}
Biography text here.
\end{IEEEbiography}






\end{document}